\documentclass{appolb}

\begin{document}
\pagestyle{plain}
\newcount\eLiNe\eLiNe=\inputlineno\advance\eLiNe by -1
\title{THERMAL EXPANSION AND MAGNETOSTRICTION STUDIES OF A KONDO LATTICE COMPOUND: CeAgSb$_{2}$
}
\author{D.T.Adroja$^{1}$, P.C. Riedi$^{2}$, J.G.M. Armitage$^{2}$ and D. Fort$^{3}$
\address{$^{1}$ISIS Facility, Rutherford Appleton Laboratory, Chilton, Didcot, Oxon, Ox11 0QX,UK. $\linebreak$$^{2}$J.F. Allen Research Laboratories, School of Physics and Astronomy, University of St. Andrews, KY16 9SS, Scotland, UK
$\linebreak$$^{3}$School of Metallurgy and Materials, University of Birmingham, B15 2TT, UK
}}
\maketitle

\begin{abstract}
We have investigated a single crystal of CeAgSb$_{2}$ using low field ac-susceptibility, thermal expansion and magnetostriction measurements. The thermal expansion coefficient $\alpha$, exhibits highly anisotropic behaviour between 3K and 80K: $\alpha$ (for $\Delta$L/L) $\perp$c exhibits a sharp peak at T$_{N}$ followed by a broad maximum at 20K, while a sharp negative peak at T$_{N}$  followed by a minimum at 20K has been observed for ($\Delta$L/L $\parallel$) the c direction. The observed maximum and minimum in $\alpha$(T) at 20K have been attributed to the crystalline field effect (CEF). The magnetostriction (MS) also exhibits anisotropic behaviour with a large MS along the c-axis. 
\end{abstract}

Recent studies on RAgSb$_{2}$ (R=rare earth) compounds have shown that these compounds crystallize in the tetragonal ZrCuSi$_{2}$ type structure [1-6]. Among these compounds, CeAgSb$_{2}$ is the most interesting. The resistivity and thermoelectric power of CeAgSb$_{2}$ show a typical Kondo lattice behaviour [1]. The magnetization exhibits strongly anisotropic behaviour with the easy c-axis of magnetization in the magnetic ordered state (probably a complex antiferromagnetic (AFM) state with a FM component) below 10K, while in the paramagnetic state there is an easy ab-plane [1,5]. The magnetization isotherm at 2K shows that the easy magnetization direction changes from the c-axis to the ab-plane above 1T field with the saturated moment $\mu$$^{sat}$=1.1$\mu$$_{B}$/Ce-ion for B$\perp$c and 0.37$\mu$$_{B}$/Ce-ion for B$\parallel$c at 5.5T field. The neutron diffraction measurements at 2K show the presence of only a single magnetic Bragg peak, (1 0 1), with moment of 0.33$\mu$$_{B}$/Ce-ion [5]. The zero-field $\mu$SR study shows well-defined frequency oscillations with an anomalously low internal field of 53mT at the muon site, which is in agreement with the extremely low frequency (0.25MG) observed in a Shubnikov-de Haas study [4, 6]. 

In the present work we have investigated a single crystal of CeAgSb$_{2}$ using ac-susceptibility ($\chi$$_{ac}$), thermal expansion, and magnetostriction measurements with the aim of throwing more light on the complex electronic and magnetic ground state. The single crystal of CeAgSb$_{2}$ was grown in an evacuated BN-crucible at 1350$^{o}$C by the Bridgemann method. The inductive component, $\chi$$^{'}$$_{ac}$(T) of the ac-susceptibility of CeAgSb$_{2}$ single crystal with B$_{ac}$(5G)$\parallel$c, exhibits a sharp peak at 9.7K which is due to the magnetic ordering of the Ce-moments (inset of Fig.1). $\chi$$^{'}$$_{ac}$(T) for B$_{ac}$$\perp$c also exhibits a similar peak, but the peak height is only 27\%\ of B$_{ac}$$\parallel$c. The c-axis is therefore the easy magnetization direction at low temperature, which is in agreement with the dc magnetization study [4]. It would be interesting to compare the $\chi$$^{'}$$_{ac}$(T) signal of CeAgSb$_{2}$ with that from the ferromagnetic CePdSb [7]. Well below T$_{N}$ (or T$_{C}$) $\chi$$^{'}$$_{ac}$(T) of CeAgSb$_{2}$ is very small, while that of CePdSb retains about 92\%\ of its peak value. This again implies that the magnetic ground state of CeAgSb$_{2}$ is more complicated than that of a simple AFM or FM.

Fig.1 shows the linear thermal expansion (TE=$\Delta$L/L) as a function of temperature for CeAgSb$_{2}$ single crystal parallel to c-axis (TE$\parallel$c) and perpendicular to c-axis (TE$\perp$c) along with the isostructural nonmagnetic reference polycrystalline LaAgSb$_{2}$. $\Delta$L/L of LaAgSb$_{2}$ exhibits a typical behaviour expected for the thermally excited phonons. On the other hand, $\Delta$L/L of CeAgSb$_{2}$ shows highly anisotropic behaviour, positive for TE$\perp$c and negative for TE$\parallel$c, with a sudden change at T$_{N}$ in both the directions. The magnetic contribution to the thermal expansion coefficient, $\alpha$$_{M}$(T) of CeAgSb$_{2}$ along both the directions was estimated by subtracting $\alpha$(T) of LaAgSb$_{2}$. $\alpha$$_{M}$(T) exhibits a sharp peak and a broad peak at T$_{N}$ and 20K, positive for TE$\perp$c and negative for TE$\parallel$c, respectively. It should be noted that the $\alpha$$_{M}$(T) of polycrystalline CeAgSb$_{2}$ also shows a broad peak at 18K, but no clear peak at T$_{N}$ [3]. The absence of the peak at T$_{N}$ in the polycrystalline sample might be due to the cancellation of positive (for TE$\perp$c) and negative (for TE$\parallel$c) contributions observed in the single crystal. The sharp peak at T$_{N}$ in CeAgSb$_{2}$ single crystal arises due to the development of anisotropic spin-spin correlations because of the magnetic ordering of Ce-moments. On the other hand the broad peak (maximum and minimum) at 20K in both the directions has been attributed to the CEF effect on the J=5/2 state of the Ce$^{3+}$ ion. This is consistent with our recent high resolution inelastic neutron scattering measurements on CeAgSb$_{2}$, which show two well defined crystal field excitations, at 5.1meV and 12.4meV, as expected for the tetragonal point symmetry of the Ce ion [8]. It is interesting to note that the observed anisotropic behaviour of $\alpha$$_{M}$(T) of CeAgSb$_{2}$ is very similar to that observed for CeRhIn$_{5}$ single crystal, which also has the tetragonal crystal structure [9]. The calculated $\alpha$$_{M}$(T) for CeRhIn$_{5}$ on the basis of the CEF model exhibits a maximum and minimum around 25K for [100] and [001] directions, respectively [9]. In order to investigate the effect of magnetic field on the $\alpha$$_{M}$(T) of CeAgSb$_{2}$, we have measured $\alpha$$_{M}$(T) in an applied magnetic field of 8T (Fig.2). The observed sharp peak at T$_{N}$ in zero field was almost suppressed in 8T field for both the directions. 

We estimated the value of dT$_{N}$/dP=-0.088 (K/kbar), using the Ehrenfest relation and the heat capacity data from Ref.[2], which is in good agreement with the experimentally measured value of -0.095 (K/kbar) on the polycrystalline CeAgSb$_{2}$ [3]. The negative sign of dT$_{N}$/dP indicates that CeAgSb$_{2}$ is on the right-hand side of the Doniach phase diagram [3]. 

Fig.3 shows the magnetostriction (MS) isotherms measured at various temperatures for ($\Delta$L/L)$\parallel$c and ($\Delta$L/L)$\perp$c with applied fields B$\parallel$c and B$\perp$c directions. MS exhibits highly anisotropic behaviour with the largest length change for ($\Delta$L/L)$\parallel$c. Between 14Kand 20K MS exhibits a quadratic behaviour, for all measured directions (Figs.3a-d), taht could be understood on the basis of the free energy of the system in an applied field. An interesting behaviour of MS is observed for ($\Delta$L/L)$\perp$c and B$\perp$c geometry at low temperatures (Fig.3d). At 3K MS exhibits a peak at 3.3T, which is consistent with the observed peak in the magnetoresistance measurements and has been attributed to the field induced transition to the easy ab-plane of magnetization [4]. Furthermore, with increasing temperature from 3K, the position of the peak moves to a lower field, which indicates that the smaller critical field for the field induce transition. A small hysteresis in MS was observed at 3K suggesting the presence of a FM component. This result along with the absence of the domain walls contribution at low fields in MS of CeAgSb$_{2}$ indicates that the magnetic ground state is not a simple FM, but a complex AFM.

In conclusion, thermal expansion and magnetostriction measurements of CeAgSb$_{2}$ single crystal exhibit highly anisotropic behaviour. These results indicate that the anisotropic magnetic exchange and CEF-anisotropy are playing an important role. 
\newline
\newline
[1] M. Houshiar et al, J. Magn. Magn. Mater, 140-144, 1232 (1995)
\newline
[2] Y. Muro et al, J. Alloys and Comp., 257, 23 (1997)
\newline
[3] M.J. Thornton et al, J. Phys. Cond. Matter, 10, 9485 (1998)
\newline
[4] K.D. Myers et al, J. Magn. Magn. Meter, 205, 27 (1999); K.D. Mayers et al, Phys. Rev. B60, 13371, (1999)
\newline
[5] G. Andre et el, Physica B292, 176 (2000) 
\newline
[6] J.A. Dann et al, Physica B, 289-290, 38 (2000)
\newline
[7] D.T. Adroja et al, Phys. Reb. B61, 1232 (2000)
\newline
[8] D.T. Adroja et al, unpublished
\newline
[9] T. Takeuchi et al, J. Phys. Soci. Japan, 70, 877 (2001).
\newline
\newline
Figure captions
\newline
Fig 1. Linear thermal expansion ($\Delta$L/L) versus temperature of CeAgSb$_{2}$ single crystal and LaAgSb$_{2}$ polycrystal. The inset shows temperature dependence of inductive, $\chi$$^{'}$$_{ac}$(T) component of the ac-susceptibility for B$_{ac}$$\parallel$c.
\newline
\newline
Fig.2 The magnetic contribution to the linear thermal expansion coefficient, $\alpha$$_{M}$(T) of CeAgSb$_{2}$ single crystal in zero field and 8T field, (a) TE$\perp$c and, (b) TE$\parallel$c. 
\newline
\newline
Fig.3  The magnetostriction isotherms at various temperatures of CeAgSb$_{2}$ single crystal, (a) and (b) ($\Delta$L/L)$\parallel$c and, (c) and (d) ($\Delta$L/L)$\perp$c, for both B $\parallel$c and B$\perp$c.
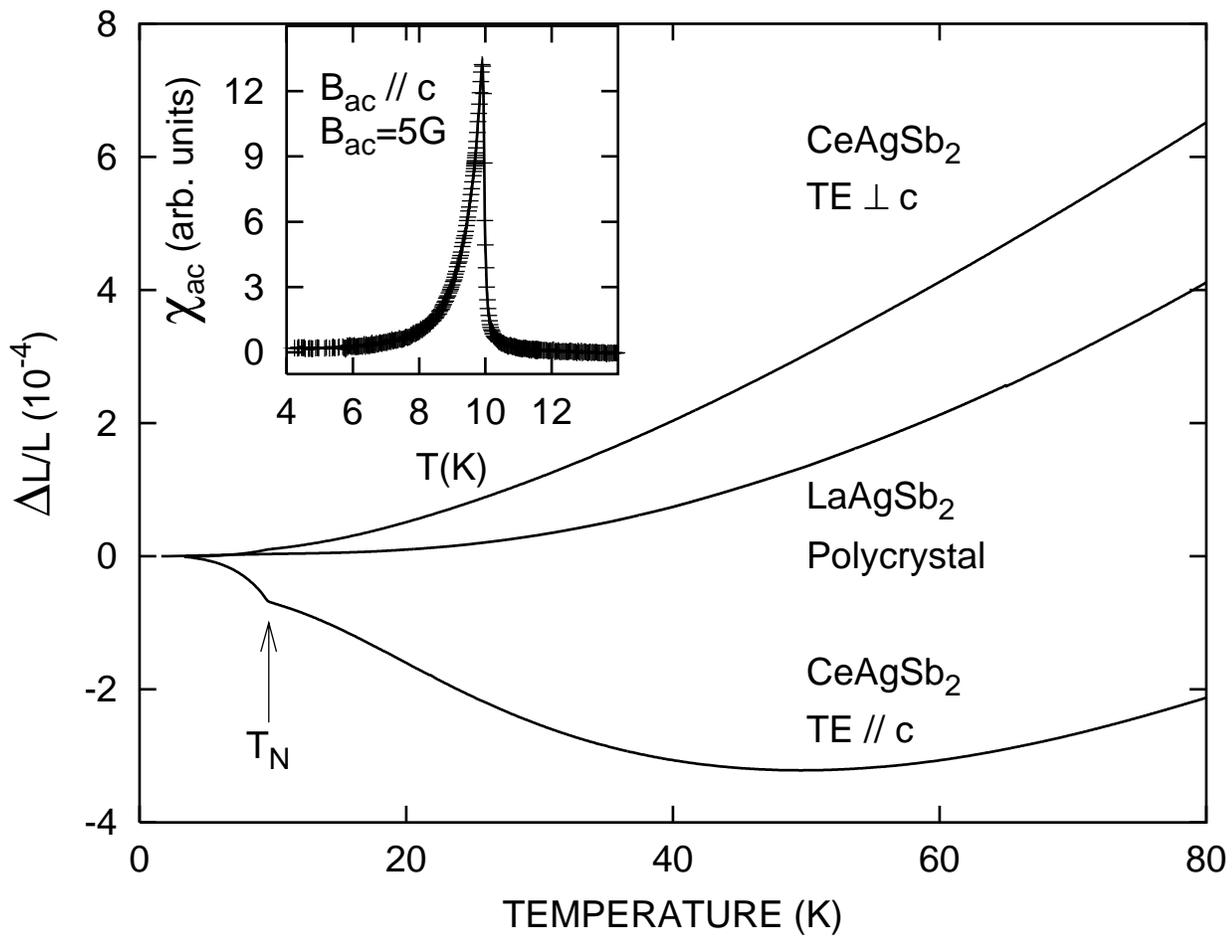
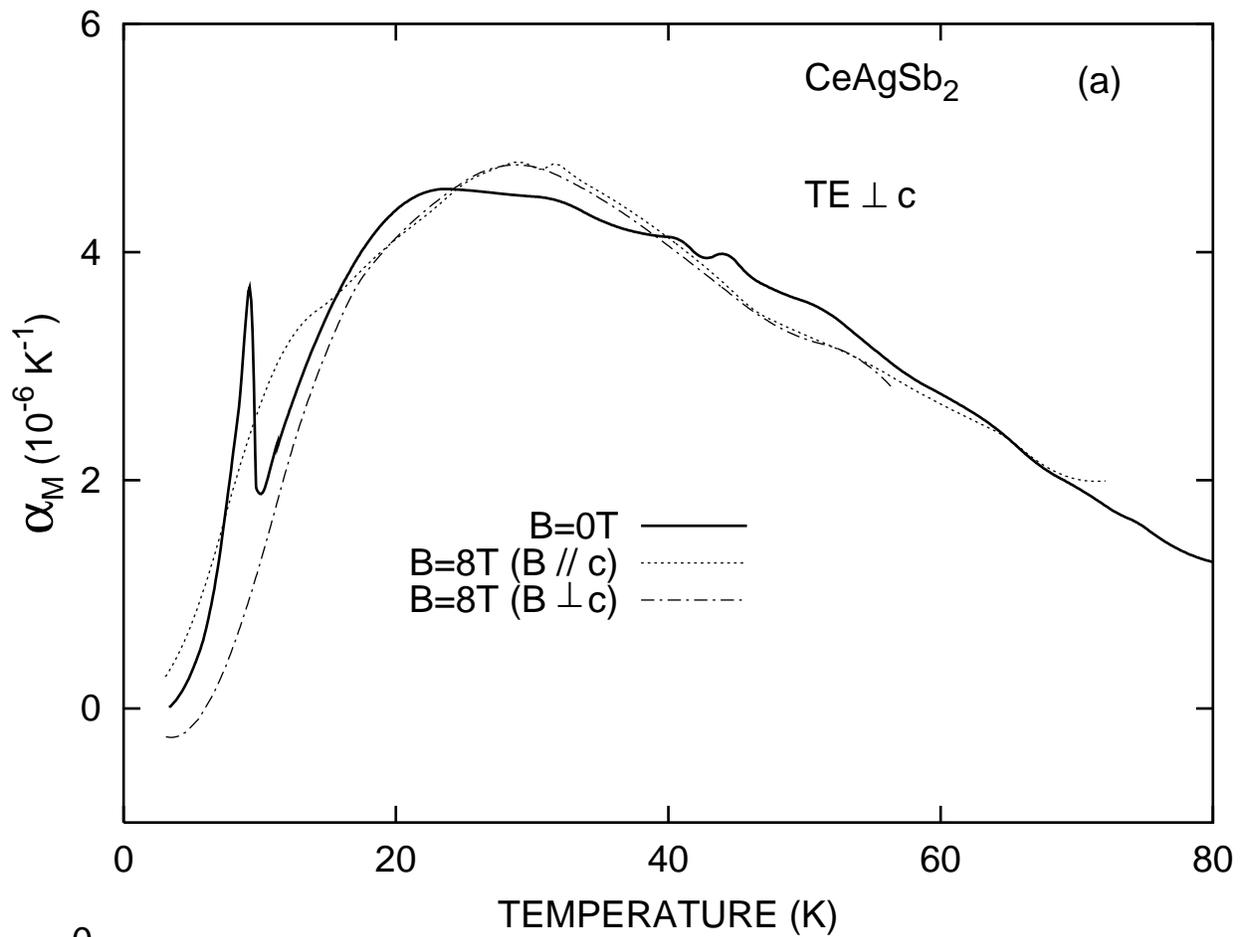
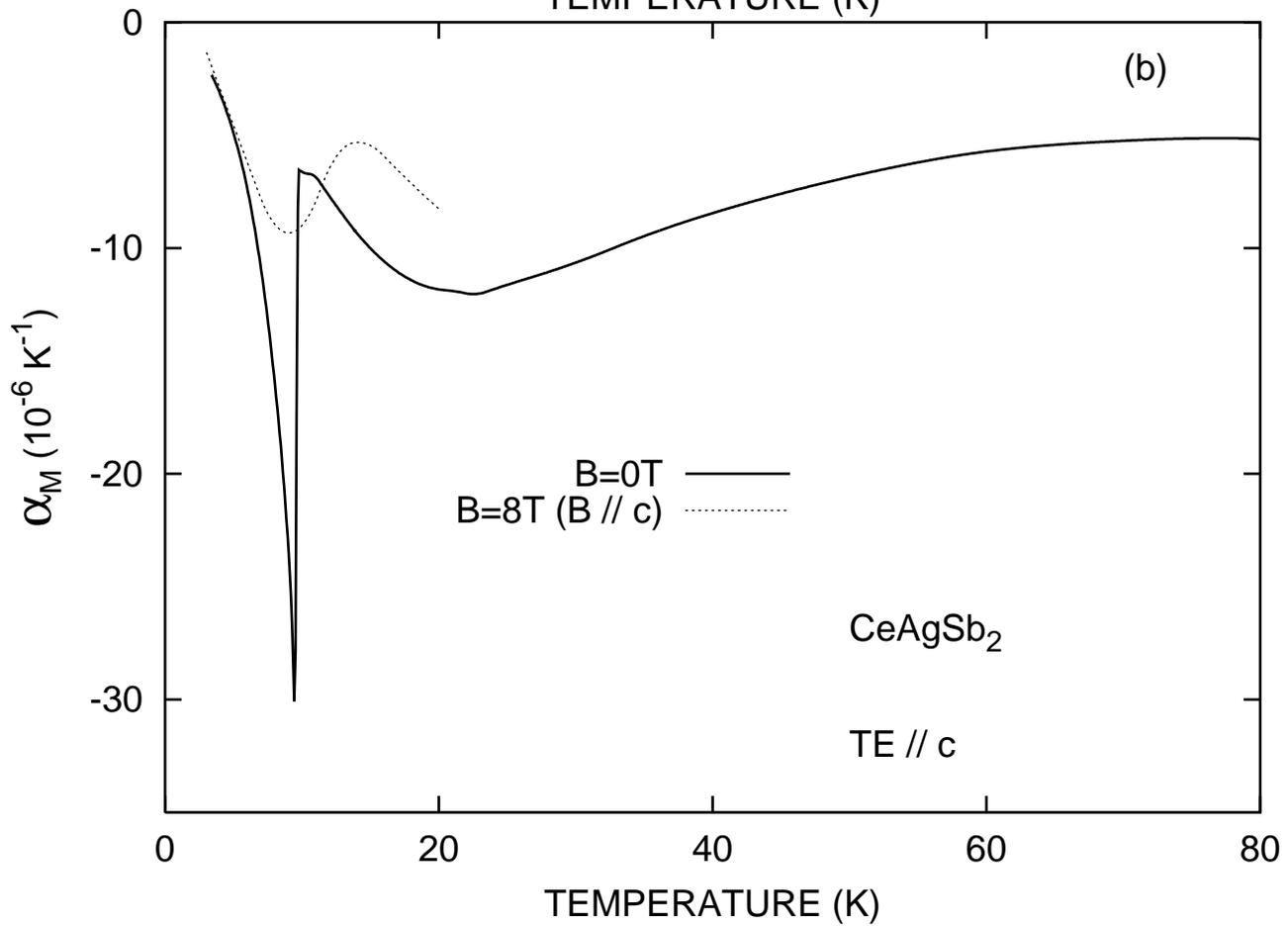
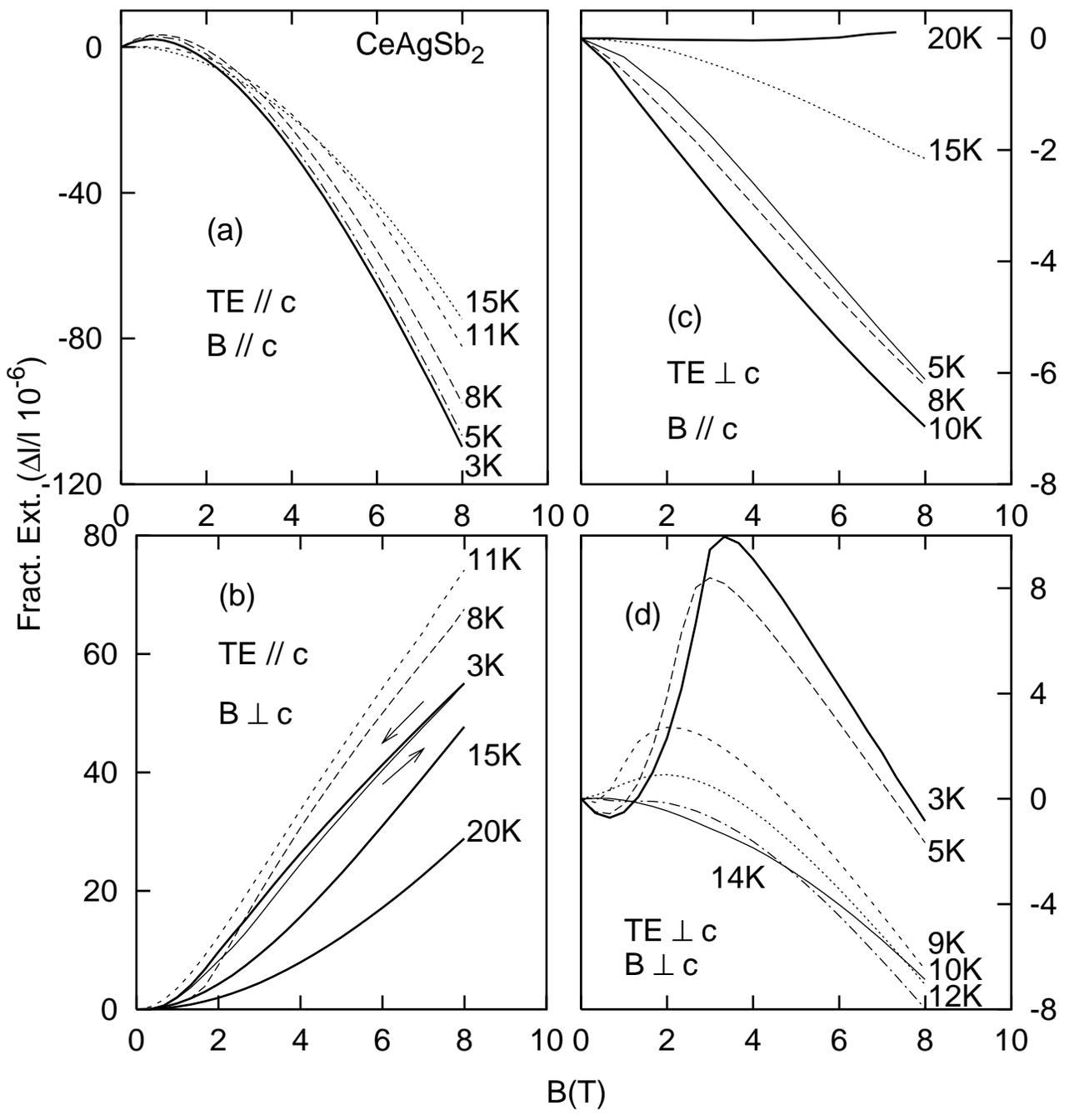
\end{document}